\documentclass[preprint,showpacs,preprintnumbers,amsmath,amssymb]{revtex4}

\usepackage{graphicx}
\usepackage{dcolumn}
\usepackage{bm}

\begin{document}


\title{Fermionic Zero Modes in Self-dual Vortex Background}

\author{Yong-Qiang Wang}
\email{wyq02@st.lzu.edu.cn}\affiliation{Institute of Theoretical
Physics, Lanzhou University, Lanzhou 730000, China}
\author{Tie-Yan Si}
\email{city02@st.lzu.edu.cn}\affiliation{Institute of Theoretical
Physics, Lanzhou University, Lanzhou 730000, China}
\author{Yu-Xiao Liu}
\thanks{Corresponding author}\email{liuyx01@st.lzu.edu.cn}\affiliation{Institute of
Theoretical Physics, Lanzhou University, Lanzhou 730000, China}
\author{Yi-Shi Duan}
\affiliation{Institute of Theoretical Physics, Lanzhou University,
Lanzhou 730000, China}
\date{\today}

\begin{abstract}

We study fermionic zero modes in the background of self-dual
vortex on a two-dimensional non-compact extra space in 5+1
dimensions. In the Abelian Higgs model, we present an unified
description of the topological and non-topological self-dual
vortex on the extra two dimensions. Based on it, we study
localization of bulk fermions on a brane with inclusion of
Yang-Mills and gravity backgrounds in six dimensions. Through two
simple cases, it is shown that the vortex background contributes a
phase shift to the fermionic zero mode, this phase is actually
origin from the Aharonov-Bohm effect.
\end{abstract}

 \maketitle

\section{Introduction}

In four dimensional space time, the interactions of fermions in a
Nielsen-Olesen vortex background have been widely analyzed in the
literature, mainly in connection with bound states at
threshold\cite{threshold}, zero modes\cite{zeromode} and
scattering solutions\cite{scattering}. Recently, Frere, Libanov
and Troitsky have shown that a single family of fermions in six
dimensions with vector-like couplings to the Standard Model (SM)
bosons gives rise to three generations of chiral Standard Model
fermions in four dimensions\cite{libanov,libanov3}. In 5+1
dimensions, Frere $et\;al$ also studied the fermionic zero modes
in the background of a vortex-like solution on an extra
two-dimensional sphere and relate them to the replication of
fermion families in the Standard Model\cite{frere}.

The topological vortex (especially Abrikosov-Nielsen-Olesen
vortex) coupled to fermions may lead to chiral fermionic zero
modes\cite{rossi}. Usually the number of the zero modes coincides
with the topological number, that is, with the magnetic flux of
the vortex. In Large Extra Dimensions (LED) models, the chiral
fermions of the Standard Model are described by the zero modes of
multi-dimensional fermions localized in the (four-dimensional)
core of a topological defect. Unlike the classical Kaluza Klein
theory where one assumes that the extra dimensions should be small
and cover a compact manifold, the extra dimensions can be large
and non compact\cite{rubakov,daemi}. This freedom can provide new
insights for a solution of gauge hierarchy problem\cite{randal}
and cosmological constant problem. While we shall study fermionic
zero modes coupled with a self-dual vortex background\cite{WangYQ}
on a two dimensional non-compact extra space in 5+1 dimensions.

The paper is organized as follows: In section II, we first present the unified
description of the topological and non-topological self-dual vortex in Abelian Higgs
model on the two dimensional non-compact extra space. In section III, In 5+1 dimensions,
we analyzed the effective lagrangian of the fermions localized on a brane in the
background of the coupling between Higgs field and fermion spinor field. In section IV,
two simple cases are discussed to show the role of vortex background in the fermionic
zero modes. In the last section, a brief conclusion is presented.

\section{Self-dual Vortex on a two-dimensional non-compact extra space}

We consider a 5+1 dimensional space-time $M^{4}\times{R^{2}}$ with
$M^{4}$ represents our four-dimensional space-time and $R^{2}$
represents the  two-dimensional extra Euclidean space. The metric
$G_{MN}$ of the manifold $M^{4}\times{R^{2}}$ is
\begin{equation}\label{metric}
ds^2 = G_{MN}dx^M dx^N = g_{\mu \nu}dx^\mu{d}x^\nu -\delta_{ij}
dx^{i}dx^{j},
\end{equation}
where $g_{\mu \nu}$ is the four dimensional metric of the manifold $M^4$, capital Latin
indices $M,N=0,\cdots,5$, Greek indices $\mu,\nu=0,\cdots,3$, lower Latin indices
$i,j=4,5$ and $x^4, x^5$ are the coordinates on $R^2$. To generate the vortex solution,
we introduce the Abelian Higgs Lagrangian
\begin{equation}\label{lagrangian}
\mathcal{L}_{V}=\sqrt{-G} \left
(-\frac{1}{4}F_{MN}F^{MN}+(D^{M}\phi)^{\dag}(D_{M}\phi)-\frac{\lambda}{2}(\|\phi\|^{2}-v^{2})^{2}
\right ),
\end{equation}
where $G=\det(G_{MN})$, $F_{MN}=\partial_{M}A_{N}-\partial_{N}A_{M}$,
$D_{M}\phi=(\partial_{M}-ieA_{M})\phi$, $\phi=\phi(x^4, x^5)$ and $A_{M}$ are a complex
scalar field on $R^2$ and a U(1) gauge field, respectively,
$\|\phi\|=(\phi\phi^{\ast})^{\frac{1}{2}}$.

The Abrikosov-Neilsen-Olsen vortex solution on the $M^{4}\times{R^{2}}$ could be
generated from the Higgs field. We first introduce the first-order Bogomol'nyi self-dual
equations \cite{Bogo} on the two dimensional extra Euclidean space
\begin{eqnarray}\label{bogo}
D_{\pm}\phi=0,\;\;B&=& \partial _{i} \partial _{i} \ln(\|\phi\|^{2})\nonumber\\
&=&\pm{e}(\|\phi\|^{2}-v^{2}), (i=4,5)
\end{eqnarray}
here the operator $D_{\pm}$ is defined as
$D_{\pm}\equiv(D_{4}\pm{i}D_{5})$. We know that complex Higgs
field $\phi$ can be regarded as the complex representation of a
two-dimensional vector field $\vec{\phi}=(\phi^{1}, \phi^{2})$
over the base space time, it is actually a section of a complex
line bundle on the base manifold. Considering the self-dual
equation $D_{+}\phi=0$ and $\phi=\phi^{1}+i\phi^{2}$, we split the
real part form the imaginary part, and obtain two equations
\begin{eqnarray}\label{split}
\left.%
\begin{array}{c}
  \partial_{4}\phi^{1}-\partial_{5}\phi^{2} = eA_{4}\phi^{2}+eA_{5}\phi^{1}, \\
  \partial_{4}\phi^{2}+\partial_{5}\phi^{1} = eA_{5}\phi^{2}-eA_{4}\phi^{1}. \\
\end{array}%
\right.
\end{eqnarray}
Substituting Eqs. (\ref{split}) and $\phi=\phi^{1}+i\phi^{2}$ into
$\partial_{4}\phi^{\ast}\phi-\partial_{4}\phi\phi^{\ast}$, it is
easy to verify
\begin{equation}\label{mu=mu+nv}
\partial_{4}\phi^{\ast}\phi-\partial_{4}\phi\phi^{\ast} = 2ieA_{4}\|\phi\|^{2} +
i(\partial_{5}\phi^{\ast}\phi+\partial_{5}\phi\phi^{\ast}).
\end{equation}
Considering the fundamental identity
\begin{equation}\label{na=phi/phi*}
\epsilon_{ab}n^{a}\partial_{i}n^{b}=\frac{1}{2i}\frac{1}{\phi\phi^{\ast}}(\partial_{i}\phi^{\ast}\phi-\partial_{i}\phi\phi^{\ast})
\end{equation}
with the unit vector defined as $n^{a}={\phi^{a}}/{\|\phi\|}, (a,b=1,2)$, we immediately
have
\begin{equation}\label{eAmu}
eA_{4}=\epsilon_{ab}n^{a}\partial_{4}n^{b}-\frac{1}{2}\partial_{5}\ln(\phi\phi^{\ast}).
\end{equation}
When considering
$\partial_{5}\phi^{\ast}\phi-\partial_{5}\phi\phi^{\ast}$,
following the same process above, it yields
\begin{equation}\label{eAnu}
eA_{5}=\epsilon_{ab}n^{a}\partial_{5}n^{b}+\frac{1}{2}\partial_{4}\ln(\phi\phi^{\ast}).
\end{equation}
Eq. (\ref{eAmu}) and Eq. (\ref{eAnu}) can be unified into one
equation
\begin{equation}\label{eAi-}
eA_{i}=\epsilon_{ab}n^{a}\partial_{i}n^{b}-\frac{1}{2}\epsilon_{i
j}\partial_{j}\ln(\phi\phi^{\ast}).
\end{equation}
Repeating the same discussion above to $D_{-}\phi=0$, we arrive
\begin{equation}\label{eAi+}
eA_{i}=\epsilon_{ab}n^{a}\partial_{i}n^{b} +
\frac{1}{2}\epsilon_{i j}\partial_{j}\ln(\phi\phi^{\ast}).
\end{equation}
In fact, since the magnetic field is $B=\epsilon^{ij}\partial_{i}(e{A}_{j})$, according
to Eq. (\ref{eAi+}), we have
\begin{equation}\label{B=delta+ln}
B=\epsilon^{i j}\epsilon_{ab}\partial_{i}n^{a}\partial_{j}n^{b}+\partial_{i}\partial_{i}\ln(\|\phi\|^{2}).\nonumber\\
\end{equation}
So the second one in the Bogomol'nyi self-dual equation (\ref{bogo}) can be generalized
to
\begin{equation}\label{nonlinear01}
B=\epsilon^{i j} \epsilon_{ab}\partial_{i}n^{a}\partial_{j}n^{b}
+\partial_{i}\partial_{i}\ln(\|\phi\|^{2})=\pm e(\|\phi\|^{2}-v^{2}).
\end{equation}
For clarity we denote
\begin{eqnarray}\label{B=Bt+Bnt}
B&=&B_{T}+B_{NT},\nonumber\\
B_{T}&=&\epsilon^{i j}\epsilon_{ab}\partial_{i}n^{a}\partial_{j}n^{b},\\
B_{NT}&=&\partial_{i}\partial_{i}\ln(\|\phi\|^{2}).\nonumber
\end{eqnarray}
According to Duan's topological current theory, it is easy to see
that the first term of Eq. (\ref{nonlinear01}) bear a topological
origin.  From Duan's $\phi$-mapping topological current theory
\cite{DuanSLAC}, one can see that the topological term of the
magnetic field $B_{T}$
\begin{equation}
B_{T}=\epsilon^{i j}\epsilon_{ab}\partial _{i}n^{a}\partial_{j}n^{b},\label{nonsigma}
\end{equation}
just describes the non-trivial distribution of $\vec{n}$ at large
distances in space \cite{'tHooft}. Noticing
$\partial_{i}n^{a}=\frac{\partial_{i}\phi^{a}}{\parallel\phi\parallel}+\phi^{a}\partial_{i}\frac{1}{\parallel\phi\parallel}$
and the Green function relation in $\phi-$space :
$\partial_{a}\partial_{a}ln(\|\phi\|)=2\pi\delta^{2}(\vec{\phi}),(\partial_{a}={\frac{\partial}{\partial\phi^{a}}})$,
it can be proved that \cite{honseng}
\begin{equation}
B_{T}=\delta^{2}({\phi})J(\frac{\phi}{x})\label{BT=del}.
\end{equation}
So the second one in the Bogomol'nyi self-dual equations (\ref{bogo}) should be
\begin{equation}\label{gnonlinear01}
B=\delta^{2}({\phi})J(\frac{\phi}{x})+\partial_{i}\partial_{i}\ln\|\phi\|^{2}=\pm
e(\|\phi\|^{2}-v^{2}).
\end{equation}
This equation is more exact than the conventional self-dual equation, in which the
topological term has been ignored all the time. Obviously when the field $\phi\neq0$, the
topological term vanishes and we have
\begin{equation}\label{B=ln}
B=B_{NT}=\partial_{i}\partial_{i}\ln(\|\phi\|^{2}).
\end{equation}
So the self-dual equation (\ref{gnonlinear01}) reduces to a nonlinear elliptic equation
for the scalar field density $\|\phi\|^{2}$
\begin{equation}\label{nonlinear}
\partial_{i}\partial_{i}\ln(\|\phi\|^{2})=\pm e(\|\phi\|^{2}-v^{2}).
\end{equation}
This is just the conventional self-dual equation. Comparing this equation with Eq.
(\ref{gnonlinear01}), one see that the topological term
$\delta^{2}(\vec{\phi})J(\frac{\phi}{x})$ is missed. The exact self-dual equation should
be Eq. (\ref{gnonlinear01}). From our previous work, obviously the first term of Eq.
(\ref{gnonlinear01}) describes the topological self-dual vortices. As for conventional
self-dual nonlinear equation (\ref{nonlinear}), a great deal of work has been done by
many physicists on it, and a vortex-like solutions was given by A. Jaffe \cite{jaffe}.
But no exact solutions are known.

Now we see that there are two classes of vortex which arise correspondingly from the
symmetric phase and asymmetric phase of the Higgs field. These two classes of vortex
provide different vortex background. And we shall study fermionic zero modes coupled with
the vortex background in the following two sections.

\section{Fermionic zero modes in the vortex background}
The lagrangian of the fermions in the vortex background on $R^{2}$
is
\begin{equation}
\mathcal{L}=\sqrt{-G} \left\{ \bar{\Psi} \Gamma^A E^M_A (\partial_M - \Omega_M +
A_M)\Psi-g\phi\Psi^{\dag}\Psi \right\}. \label{Lpsi}
\end{equation}
where $E^{M}_{A}$ is the {\sl sechsbein} with
\begin{equation}
E^{M}_{A} = (e^{\mu}_{A},\delta^{4}_{A},\delta^{5}_{A})
\end{equation}
and capital Latin indices $A,B=0,\cdots,5$ correspond to the flat
tangent six-dimensional Minkowski space, $\Omega_M=\frac{1}{2}
\Omega_M^{AB}I_{AB}$ is the spin connection with the following
representation of six-dimensional 8 $\times$ 8 Dirac matrices
$\Gamma^A$:
\begin{equation}
\Gamma^A=\begin{pmatrix}
  0 & \Sigma^A \\
  \bar{\Sigma}^A & 0 \\
\end{pmatrix}
,
\end{equation}
where $\Sigma^0 = \bar{\Sigma}^0 = \gamma^0 \gamma^0$; $\Sigma^k =
-\bar{\Sigma}^k = \gamma^0 \gamma^k (k=1,2,3)$; $\Sigma^4 = -\bar{\Sigma}^4 = i\gamma^0
\gamma^5$; $\Sigma^5 = -\bar{\Sigma}^5 = \gamma^0$, $\gamma^{\mu}$ and $\gamma^5$ are
usual four-dimensional Dirac matrices in the chiral representation:
\begin{equation} \gamma^0=\begin{pmatrix}
  0 & 1 \\
  1 & 0 \\
\end{pmatrix},\;\;\;
\gamma^k=\begin{pmatrix}
  0 & \sigma^k \\
  -\sigma^k & 0 \\
\end{pmatrix},\;\;\;
\gamma^5=i \gamma^0\gamma^1\gamma^2\gamma^3=\begin{pmatrix}
  1 & 0 \\
  0 & -1 \\
\end{pmatrix},
\end{equation}
here $\sigma^k$ are the Pauli matrices. $\Gamma^A$ follows relation
$\Gamma^A\Gamma^B+\Gamma^B\Gamma^A=2\eta^{AB}I$, in which $\eta^{AB}=$ diag$
(+,-,\cdots,-)$ is the six-dimensional Minkowski metric.

The components of $\Omega_M$ are
\begin{equation}\label{connection}
\Omega_{\mu}=\omega_{\mu},\;\;\; \Omega_{4}=0, \;\;\;\
\Omega_{5}=0,
\end{equation}
\noindent where $\omega_{\mu}=\frac{1}{2} \omega_{\mu}^{ab}I_{ab}$
is the spin connection derived from the metric
$g_{\mu\nu}=e_{\mu}^{a} e_{\nu}^{b} \eta_{ab}$, lower case Latin
indices $a,b=0,\cdots,3$ correspond to the flat tangent
four-dimensional Minkowski space.

Using Eqs. (\ref{connection}), the lagrangian (\ref{Lpsi}) of the fermions then becomes
\begin{eqnarray}\label{Lpsi2}
\mathcal{L}=\sqrt{-G} \{ \bar{\Psi} \Gamma^a e^{\mu}_{a} (\partial_{\mu} - \omega_{\mu} +
A_{\mu})\Psi + \bar{\Psi}\Gamma^4(\partial_{4}+A_{4})\Psi
 + \bar{\Psi} \Gamma^5(\partial_{5} + A_{5})\Psi - g
\phi \Psi ^{\dag} \Psi \}.
\end{eqnarray}
We denote the Dirac operator on $R^2$ with $D_R$:
\begin{equation} \label{DiracOperator}
D_R =\bar{\Gamma} \left \{ \Gamma^4(\partial_{4} +  A_{4}) +
\Gamma^5(\partial_{5} + A_{5})-g\phi \right \},
\end{equation}
where $\bar{\Gamma}=\Gamma^0 \Gamma ^1 \Gamma ^2 \Gamma ^3$, and expand any spinor
$\Psi(x^{\mu},x^i)$ in a set of eigenvectors $\Theta_m(x^i)$ of this operator $D_R$
\begin{equation}
D_R\Theta_m(x^i)=\lambda_m\Theta_m(x^i) \;\;\; (i=4,5).
\end{equation}
There may exist a set of discrete eigenvalues $\lambda_m$ with
some separation. All these eigenvalues play a role of the mass of
the corresponding four-dimensional excitations \cite{libanov}. We
assume that the energy scales probed by a four-dimensional
observer are smaller than the separation, and thus even the first
non-zero level is not excited. So, we are interested only in the
zero modes of $D_R$:
\begin{equation} \label{DiracEqOnS}
D_R\Theta(x^i)=0.
\end{equation}
This is just the Dirac equation on $R^2$ with gauge and vortex
backgrounds. For fermionic zero modes, we can write
\begin{equation}
\Psi(x^{\mu},x^i)=\psi(x)\Theta(x^i),
\end{equation}
where $\psi$ and $\Theta$ satisfy
\begin{eqnarray}
\bar{\Gamma}\Gamma^a e^{\mu}_{a} (\partial_{\mu} - \omega_{\mu} + A_{\mu})\psi(x) &=& 0, \nonumber\\
D_R\Theta(x^i) &=& 0.
\end{eqnarray}
The effective Lagrangian for $\psi$ then becomes
\begin{eqnarray}
&&\int dx^4 dx^5 \sqrt{-G}
\left \{ \bar{\Psi} \Gamma^A E^M_A (\partial_M - \Omega_M + A_M) \Psi\ -g\phi\Psi^{\dag}\Psi \right\}  \nonumber\\
&&= \sqrt{-\det(g_{\mu\nu})} \bar{\psi} \Gamma^a e^{\mu}_a (\partial_{\mu} - \omega_{\mu}
+ A_{\mu}) \psi \int dx^4 dx^5 \Theta^\dag \Theta.
\end{eqnarray}
Thus, to have the localization of gravity and finite kinetic energy for $\psi$, the above
integral must be finite. This may be achieved for some $\Theta(x^i)$ which does not
diverge on the whole $R^2$ and converge to zero as $r$ tends to infinity.

\section{SIMPLE SITUATIONS}
In this section, to illustrate how the vortex background affects the fermionic zero mode,
we first discuss the simple case that the Higgs field $\phi$ is relative only to $x^4$,
and then solve the general Dirac equation for the vacuum Higgs field solution
$\|\phi\|^{2}=v^2$.

\subsection{Case I: $\phi$ is relative only to $x^4$}
Now we discuss a simple situation for $\phi$, i.e.
$\phi=\phi(x^4)=\phi^1(x^4)+i \phi^2(x^4)$. In this case, Eq.
(\ref{eAmu}) and Eq. (\ref{eAnu}) can be written as
\begin{equation}\label{Atheta}
A_{4}=\frac{1}{e\|\phi\|^2} \left
(\epsilon_{ab}\phi^{a}\partial_{4}\phi^{b}
-\frac{1}{\|\phi\|^2}\epsilon_{ab}\phi^{a}\phi^{b}\phi^{c}\partial_{4}\phi^{c}
\right ),
\end{equation}
\begin{equation}\label{Aphi}
A_{5}=-\frac{1}{2e}\partial_{4}\ln \|\phi\|^{2}.
\end{equation}
Then the Dirac operator $D_R$ becomes:
\begin{equation} \label{DiracOperator2}
D_R =\bar{\Gamma} \left \{ \Gamma^4 \left[
\partial_{4} + A_{4}(x^4)- \Gamma^4 \Gamma^5 A_{5}(x^4) +\Gamma^4 g \phi(x^4) \right] +
\Gamma^5\partial_{5} \right \},
\end{equation}
and Dirac equation $D_R\Theta(x^4,x^5)=0$ is
\begin{equation} \label{DiracEqI}
 \left \{\bar{\Gamma} \Gamma^4 \left[
\partial_{4} + A_{4}(x^4)- \Gamma^4 \Gamma^5 A_{5}(x^4) +\Gamma^4 g \phi(x^4) \right ]
+ \bar{\Gamma} \Gamma^5 \partial_{5}\right \} \Theta(x^4,x^5)=0.
\end{equation}
Here $\Theta(x^4,x^5)$can be written as the following form:
\begin{equation}
\Theta(x^4,x^5)=f(x^4)h(x^5),
\end{equation}
where $h(x^5)=Const$  and $f(x^4)$ satisfies
\begin{equation} \label{DiracEqf}
\left \{ \partial_{4} + A_{4}(x^4)- \Gamma^4 \Gamma^5 A_{5}(x^4)
+\Gamma^4 g \phi(x^4) \right \} f(x^4)=0.
\end{equation}
Solving this equation, one can easily obtain the formalized solution:
\begin{eqnarray} \label{fx4_1}
f(x^4)=e^{-\int dx^4 \left \{ A_{4}(x^4)- i \gamma^5 A_{5}(x^4) + g \Gamma^4 \phi(x^4)
\right\}}.
\end{eqnarray}
This equation spontaneously leads to the Aharonov-Bohm phase. Considering
$\phi=\phi(x^4)$ and integrating over the extra dimensions for the Eq.
(\ref{gnonlinear01}), one can get
\begin{equation} \label{WindingNumber}
\partial_{4}\ln\|\phi\|^{2}=-\sum_{l}{W_{l}} \pm \int dx^4
(\|\phi\|^{2}-v^2),
\end{equation}
where $W_l$ is winding number. Making use of Eq. (\ref{Aphi}) and substituting Eq.
(\ref{WindingNumber}) into Eq. (\ref{fx4_1}), we get the following form
\begin{eqnarray}\label{fx4WindingNumber}
f(x^4)=e^{\frac{i\gamma^5}{2 e }  \sum_{l}{W_{l}} -\int dx^4 \left \{ A_{4}(x^4)
 \pm\frac{i\gamma^5}{2 e}   \int dx^4
(\|\phi\|^{2} - v^2)  + g \Gamma^4 \phi(x^4) \right\}}.
\end{eqnarray}
From the first term $e^{\frac{i}{2e}\gamma^5\sum_{l}{W_{l}}}$, we see that the total
topological charge $Q=\sum{W_{l}}$ contributes a phase factor to the zero mode
$\Theta(x^i)=Cf(x^4)$. The topological charge is determined by the topological properties
of the extra space manifold. When we add a point in the infinity, the non-compact $R^{2}$
can be compactified to a 2-sphere. In this case, the total topological charge is just the
Euler characteristic number of 2-shpere, i.e., $Q=2$. Eq. (\ref{gnonlinear01}) reveals
that this topological phase origin from the symmetric phase of the Higgs field, and the
non-topological one arise from asymmetric phase, it is also included in
$e^{\pm\frac{i}{2e}\gamma^5\int dx^4(\|\phi\|^{2}-v^2)}$. So the topological and
non-topological self-dual vortex both contribute a phase shift to the fermionic zero
mode.

As all known, quantum topological and geometrical phases are
ubiquitous in modern physics---in cosmology, particle physics,
modern string theory and condensed matter. In fact, according to
Eq. (\ref{fx4WindingNumber}), we see this phase shift is actually
the quantum mechanical Aharonov-Bohm phase. This discission can be
generalized to the AB phase of non-abelian gauge theories, such as
the Wilson and 't Hooft loops. Since the AB phase is fundamental
to theories of anyons and to gauge fields, it is an important tool
for studying the issues of confinement and spontaneous symmetry
breaking.

\subsection{Case II: The vacuum solution}

For the vacuum solution of Eq. (\ref{gnonlinear01}) $\|\phi\|^{2}=\phi\phi^{\ast}=v^2$
which represents a circle $S^{1}$ in the extra space, according to Eq. (\ref{eAi+}), we
see the non-topological part $\frac{1}{2}\epsilon_{i j}\partial_{j}\ln(\phi\phi^{\ast})$
vanishes, there is only topological part left. When the Higgs field is degenerated on the
vacuum manifold, we have $A_{4}=A_{5}=0$, then the Dirac equation $D_R \Theta(x^4,x^5)=0$
is read as
\begin{equation} \label{DiracEqII}
\left \{\bar{\Gamma} \Gamma^4  \left (
\partial_{4} +g v \Gamma^4  \right ) +
\bar{\Gamma}\Gamma^5\partial_{5} \right \} \Theta(x^4,x^5)=0.
\end{equation}
In which $\Theta(x^4,x^5)=f(x^4)h(x^5)$, $h(x^5)$ is a constant
again and $f(x^4)$ satisfy the following equation
\begin{equation}
\bar{\Gamma} \Gamma^4  \left (
\partial_{4} + g v\Gamma^4  \right ) f(x^4)=0.
\end{equation}
Denoting
\begin{equation}
f(x^4)=\left(
\begin{array}{c}
  f_1(x^4) \\
  f_2(x^4) \\
  f_3(x^4) \\
  f_4(x^4) \\
\end{array}
\right),
\end{equation}
one obtains the following two sets of the differential equations
\begin{equation}
\left \{
\begin{array}{c}
\partial_{4} f_1(x^4) -i g v f_4(x^4) = 0, \\
\partial_{4} f_4(x^4) -i g v f_1(x^4) = 0; \\
\end{array}
\right.
\end{equation}
\begin{equation}
\left \{
\begin{array}{c}
\partial_{4} f_2(x^4) +i g v f_3(x^4) = 0, \\
\partial_{4} f_3(x^4) +i g v f_2(x^4) = 0. \\
\end{array}
\right.
\end{equation}
The solutions are
\begin{eqnarray}
\left.
\begin{array}{l}
f_1(x^4) = C_1 e^{i Q x^4} + C_4 e^{- i Q x^4},\\
f_2(x^4) = C_2 e^{ Q x^4} + C_3 e^{- Q x^4},\\
f_3(x^4) = i(C_2 e^{ Q x^4} - C_3 e^{- Q x^4}),\\
f_4(x^4) = C_1 e^{i Q x^4} - C_4 e^{- i Q x^4},\\
\end{array}
\right.
\end{eqnarray}
where $Q=gv$. Now we see that $f_1(x^4)$ and $f_4(x^4)$ are planar wave function. It is
easy to see that, if the coupling constant $g=0$ or the vacuum expectation $v=0$, the
solution $f(x^{4})$ is simply a constant spinor. As shown in section II, $\phi=v=0$
corresponds to the symmetric phase and $\phi=v\neq0$ corresponds to the asymmetric phase.
So different vortex background results in different zero mode.

The discussion above can also be generalized to a more universal case, usually the
general Dirac equation is hardly solvable, while the two simple cases above provide us a
coarse insight into the fermionic zero modes in the vortex background.

\section{Conclusion}
In 5+1 dimensions, there are two classes of vortex solutions in the Abelian Higgs model:
the topological vortex and the non-topological vortex. They can be described by a more
accurate Bogomol'nyi self-duality equation
$B=\delta^{2}(\vec{\phi})J(\frac{\phi}{x})\pm\partial_{i}
\partial_{i}ln(\|\phi\|^{2})$. The topological vortex just arise from the
symmetric phase of the Higgs field, while the non-topological vortex origin from the
asymmetric phase. Through a simple case, it is shown that the vortex background
contribute a phase shift to the fermionic zero mode in the 5+1 dimensional space time.
The phase is divided into two parts, one is related with the topological number of the
extra space, the other depends on the non-topological vortex solution. Then we solve the
general Dirac equation for the vacuum case, the symmetric and asymmetric phases of the
Higgs field just correspond to different fermion solutions.

\section{Acknowledgment}
This work was supported by the National Natural Science Foundation
and the Doctor Education Fund of Educational Department of the
People's Republic of China.


\begin{thebibliography}{99}

\bibitem{threshold} C. R. Nohl, Phys. Rev. D \textbf{12}, 1840
(1975).

\bibitem{zeromode} R. Jackiw,  Phys. Rev. D \textbf{29}, 2375
(1984).

\bibitem{scattering} H. J. de Vega, Phys. Rev. D \textbf{18},
2932 (1978).

\bibitem{libanov} M. V. Libanov and S. V. Troitsky, Nucl. Phys. B
\textbf{599}, 319 (2001).

\bibitem{libanov3}J. M. Frere, M. V. Libanov and S. V. Troitsky,
Phys. Lett. B \textbf{512}, 169 (2001); JHEP \textbf{0111}, 025 (2001).

\bibitem{frere} J. M. Frere, M. V. Libanov, E. Ya. Nugaev and
S. V. Troitsky, JHEP \textbf{0306}, 009 (2003).

\bibitem{rossi} R. Jackiw and P. Rossi, Nucl. Phys. B
\textbf{190}, 681 (1981).

\bibitem{rubakov} V. Rubakov and M. Shaposhnikov, Phys. Lett. B
\textbf{125}, 139 (1983).

\bibitem{daemi} S. Randjbar-Daemi and C. Wetterich, Phys. Lett. B
\textbf{166}, 65 (1986).

\bibitem{randal} L. Randal and R. Sundrum, Phys. Rev. Lett.
\textbf{83}, 4690 (1997).

\bibitem{WangYQ} Y. Q. Wang, Master thesis, Lanzhou University
(2005).

\bibitem{Bogo} E. Bogomol'nyi, Sov. J.
Nucl. Phys. \textbf{24}, 449 (1976).

\bibitem{DuanSLAC} Y. S. Duan, SLAC-PUB-3301 (1984); Y. S. Duan,
L. B. Fu and G. Jia, J. Math. Phys. \textbf{41}, 4379 (2000).

\bibitem{'tHooft} G. 't Hooft, Nucl. Phys. B \textbf{79}, 276
(1974); A. Polyakov, JETP Lett. \textbf{20}, 194 (1974).

\bibitem{honseng} Y. S. Duan, S. Li and G. H. Yang, Nucl. Phys. B
\textbf{514}, 705 (1998).

\bibitem{jaffe} A. Jaffe and C. Taubes, Vortices and Monopoles,
(Birkhauser 1980).



\end{thebibliography}
\end{document}